\documentstyle[preprint,aps,epsf]{revtex}

\newcommand{\beq}{\begin{equation}}
\newcommand{\eeq}{\end{equation}}
\def\bea{\begin{eqnarray}}  \def\eea{\end{eqnarray}}

\begin{document}

\begin{center}
\vspace*{1 truecm}
{\Large \bf Transverse momentum fluctuations and 
percolation of strings
} \\[8mm]
{\bf  E. G. Ferreiro, F. del Moral and C. Pajares}\par
{\it Departamento de
F\'{\i}sica de  Part\'{\i}culas,
Universidade de Santiago de
Compostela,\\ 15782--Santiago de Compostela, Spain}

\vskip 1.0truecm
{\Large {\bf Abstract}}
\end{center}
\begin{quotation}
The behaviour of the transverse momentum fluctuations with the
centrality of the collision shown by the  
Relativistic
Heavy Ion Collider
data is naturally explained by
the clustering of color sources. In this framework, 
elementary color sources --{\it strings}-- overlap forming 
clusters, so the number of effective sources is modified.
These clusters decay into particles with mean transverse momentum that depends 
on the number of elementary sources that conform each cluster, and the area occupied by the cluster.
The transverse momentum fluctuations in this approach correspond
to the fluctuations of the transverse momentum of these clusters, and they
behave essentially as the number of effective
sources.

\end{quotation}
\vskip 1.0truecm


\vspace{5cm}

\newpage

Event-by-event fluctuations of the mean transverse momentum are considered to
be one of the most important tools to identify a phase transition in the evolution
of the system created in relativistic heavy ion collisions, since a
second order phase transition may lead to a divergence of the specific heat
which could be observed as fluctuations 
in mean transverse momentum \cite{1}-\cite{3}.
These fluctuations have been extensively studied both 
theoretically \cite{4}-\cite{10} and
experimentally \cite{11}-\cite{16}. Recently, the PHENIX collaboration \cite{15}-\cite{15b}
 of the Relativistic
Heavy Ion Collider (RHIC) has reported a peculiar behaviour of the variable
$F_{p_T}$ that measures the transverse momentum fluctuations as a function of
the number of participants in Au-Au collisions.
$F_{p_T}$ quantifies the deviation of the observed fluctuations from
statistically independent particle emission, 
\beq
F_{p_T} = \frac{\omega_{data} - \omega_{random}}{\omega_{random}}
\eeq
where
\beq
\omega= \frac{\sqrt{<p_T^2>-<p_T>^2}}{<p_T>}\ ,
\eeq
and $<p_T>$ is the mean transverse momentum averaged over all particles and all
events.
The data show that $F_{p_T}$ increases with the number of participants,
reaching a maximum around 
$N_{part}=150 \div 200$ 
and decreasing at higher centrality.
In this paper we show that this behaviour is naturally explained 
by the clustering of elementary color sources
that may take place in heavy ion collisions.

Multiparticle production is currently described in terms of color strings
stretched between the partons of the projectile and the target. These strings
decay into new ones by sea $q-\bar{q}$ production, and subsequently hadronize
to produce the observed hadrons. 
In our approach, the strings are equivalent to effective color sources with a
fixed tranverse size
$\pi r_0^2$, with $r_0 \simeq 0.2$ fm, filled with the
color field created by the colliding partons. With increasing energy and/or
atomic number of the colliding nuclei, the number of exchanged strings
grows, so they start to interact forming clusters.
In the transverse space that means that the transverse areas of the strings
overlap, as it happens for disks in the two dimensional percolation theory. 
Moreover, at a
certain critical density of disks, $\eta \simeq 1.12-1.18$, a macroscopical cluster appears
which marks the percolation phase transition \cite{17}-\cite{18},
which is a second order, non thermal, phase transition.

In the case of a nuclear collision, the density of disks --elementary strings-- corresponds to 
\beq
\eta_c=\frac{N_s S_1}{S_A}
\eeq
where $N_s$ is the total number of strings created in the collision, each one
of an area $S_1=\pi r_0^2$, and $S_A$ corresponds to the nuclear overlap area,
$S_A=\pi R^2_A$ for central collisions.

The percolation theory governs the geometrical pattern of the string
clustering. Its observable implications, however, required the introduction of
some dynamics in order to describe the behaviour of the cluster formed by
several overlapping strings \cite{19}-\cite{20}. 
We assume that a cluster of $n$ strings that occupies an area $S_n$
behaves as a single color source with a higher color field, generated
by a higher color charge
$Q_n$. This charge
corresponds to the vectorial sum of the color charges of each individual
string ${\bf Q}_1$. The resulting color field covers the area $S_n$ of
the cluster. As $Q_n^2=( \sum_1^n {\bf Q}_1)^2$,
and the individual
string colors may be oriented in an arbitrary manner respective to one
another, the average ${\bf Q}_{1i}{\bf Q}_{1j}$ is zero, so
$Q_n^2=n  Q_1^2$.
$Q_n$ depends also on the area $S_1$ of each individual string
that comes into the cluster, as well as on
the total area of the cluster $S_n$, $Q_{n}= \sqrt{ n S_n \over S_1} Q_{1}$ \footnote{
$Q_n$ would be equal to $\sqrt{ n} Q_1$ if the strings overlap completely. Since the strings may 
overlap only partially we introduce a dependence on the area of the cluster. See first Ref. of 
{\protect \cite{19}} for more
details.}.
We take $S_1$
constant and equal to a disk of radius $r_0 \simeq 0.2$ fm.
$S_n$ corresponds to the total area occupied by
$n$ disks \cite{19}.
One could do reasonable alternative assumptions about the interaction among
the strings, but they have incompatibilities with correlation data
\cite{23}-\cite{24}.

Notice that if the strings are just touching each other, $S_n=n S_1$
and $Q_{n}=n Q_{1}$,
so
the strings behave independently. On the contrary, if they
fully overlap,
$S_n=S_1$ and $Q_{n}=\sqrt {n}Q_{1}$.
Knowing the color charge $Q_n$, one can compute the multiplicity $\mu_n$ and
the mean transverse momentum $<p_T>_n$ of the particles produced by a cluster
of $n$ strings.
According to the Schwinger mechanism for the fragmentation of the cluster, one
finds
\beq
\mu_n=\sqrt{\frac{n S_n}{S_1}} \mu_1\ {\rm and}\ <p_T>_n=\Big ( \frac{n S_1}{S_n}
\Big )^{1/4} <p_T>_1
\eeq
for the multiplicity $\mu_n$ and the average transverse momentum $<p_T>_n$
of the particles produced by a cluster
formed by $n$ strings, where $\mu_1$ and $<p_T>_1$ correspond to
the mean multiplicity and the mean transverse momentum of the particles
produced by one individual string.
 These equations constitute the main tool of
our evaluations.

The behaviour of the transverse momentum fluctuations can be understood as
follows:
At low density, most of the particles are produced by individual strings with
the same $<p_T>_1$, so the fluctuations are small. Similarly, at large density
above the percolation critical point, there is essentially only one
cluster formed by most of the strings created in the collision and therefore
fluctuations are not expected either. Instead, the fluctuations are expected to
be maximal 
below the percolation critical density, where 
the number of clusters is larger. 
Moreover, there are clusters
formed by very different numbers of strings, 
with different size,
and therefore with different
$<p_T>_n$.

In order to develop quantitatively this idea, we introduce the function \cite{4}
$\phi$ defined by
\beq
\phi=\sqrt{\frac{<Z^2>}{<\mu>}}-\sqrt{<z^2>}\ .
\eeq
$F_{p_T}$ is related to $\phi$ \cite{14}, approximately
\beq
F_{p_T}=\frac{\phi}{\sqrt{<z^2>}}=\frac{1}{\sqrt{<z^2>}}
\sqrt{\frac{<Z^2>}{<\mu>}} -1\ .
\eeq
For each particle we define $z_i={p_T}_i - <p_T>$, where ${p_T}_i$ is the
transverse momentum of the particle $i$ and $<p_T>$ is the mean transverse
momentum of all particles averaged over all events.
$\sqrt{<z^2>}$ is the second moment of the single particle inclusive $z$
distribution, and it is averaged over all events. 
$Z$ is defined 
for each event,
\beq
Z_i=\sum_{j=1}^{N_i} z_j
\eeq
where $N_i$ is the number of particles produced in an event $i$. 

In this way, introducing our formulae for the multiplicity and the mean $p_T$
we get:
\beq
<p_T>=\frac{\sum_{i=1}^{N_{events}} \sum_j \mu_{{n_j}} <p_T>_{{n_j}}}
{\sum_{i=1}^{N_{events}} \sum_j \mu_{{n_j}}} \ .
\eeq
The sum over $j$ goes over all individual clusters $j$, each one formed by ${{n_j}}$
strings and occupying an area $S_{{n_j}}$.  
The quantities ${{n_j}}$ and $S_{{n_j}}$
are obtained for each event, using a Monte Carlo code \cite{21}-\cite{22},
based on the quark gluon string model. Each string is generated at an
identified impact parameter in the transverse space. Knowing the tranverse area
of each string, we identified all the clusters formed in each event, the
number of strings ${{n_j}}$ that conforms each cluster $j$, and the area  
occupied by each cluster $S_{{n_j}}$. Note that for two different clusters, $j$ and $k$, formed by the
same number of strings ${{n_j}}=n_k$, the areas $S_{{n_j}}$ and $S_{n_k}$ can vary.
Because of this we do the sum over all individual clusters.

For the quantities $<p_T>_{{n_j}}$ --mean $p_T$ of the particles produced by 
a cluster $j$ of ${n_j}$ strings and area $S_{{n_j}}$--  and $\mu_{{n_j}}$ --mean
multiplicity of a cluster $j$ formed by
${n_j}$
strings and of area $S_{{n_j}}$--
we apply the analytical expressions given by eqs. (4).
Finally we do the average over all events.

By introducing eqs. (4) for $<p_T>_{{n_j}}$ and $\mu_{{n_j}}$ we get:
\bea
<p_T>=\frac{\sum_{i=1}^{N_{events}} \sum_j \Big ( \frac{{n_j} S_{n_j}}{S_1}
\Big )^{1/2} \mu_1 \Big ( \frac{{n_j} S_1}{S_{n_j}} \Big )^{1/4} <p_T>_1}
{\sum_{i=1}^{N_{events}} \sum_j \Big ( \frac{{n_j} S_{n_j}}{S_1}
\Big )^{1/2} \mu_1} \nonumber \\
=\frac{\Big < \sum_j {n_j}^{3/4} (\frac{S_{n_j}}{S_1})^{1/4} \Big >}
{\Big < \sum_j  (\frac{{n_j} S_{n_j}}{S_1})^{1/2} \Big >}
<p_T>_1 \nonumber \\
 =
f_2 <p_T>_1
\eea
where the mean value in the r.h.s. corresponds to an average over all events.

For the quantities $<z^2>$ and $<Z^2>$ we obtain:
\bea
<z^2>=\frac{\sum_{i=1}^{N_{events}} \sum_j \Big ( \frac{{n_j} S_{n_j}}{S_1}
\Big )^{1/2} \mu_1 \Big [ \Big ( \frac{{n_j} S_1}{S_{n_j}} \Big )^{1/4} <p_T>_1 -
<p_T>\Big ]^2}
{\sum_{i=1}^{N_{events}} \sum_j \Big ( \frac{{n_j} S_{n_j}}{S_1}
\Big )^{1/2} \mu_1} \nonumber \\
=\Big [ \frac{\Big < \sum_j {n_j} \Big >}{\Big < \sum_j  (\frac{{n_j}
S_{n_j}}{S_1})^{1/2} \Big >} - f_2^2 \Big ] <p_T>_1^2 \nonumber \\
=(f_1-f_2^2) <p_T>_1^2
\eea
and
\bea
\frac{<Z^2>}{<\mu>}= \frac{\sum_{i=1}^{N_{events}} \Big [ \sum_j \Big
( \frac{{n_j} S_{n_j}}{S_1}
\Big )^{1/2} \mu_1 \Big [ \Big ( \frac{{n_j} S_1}{S_{n_j}} \Big )^{1/4} <p_T>_1 -
<p_T> \Big ] \Big ]^2 }
{\sum_{i=1}^{N_{events}} \sum_j \Big ( \frac{{n_j} S_{n_j}}{S_1}
\Big )^{1/2} \mu_1} \nonumber \\
= \Big [ \frac{\Big < \Big [ \sum_j {n_j}^{3/4} \Big ( \frac{S_{n_j}}{S_1} \Big )^{1/4}\Big ]^2 \Big >}
{\Big < \sum_j \Big ( \frac{{n_j} S_{n_j}}{S_1}
\Big )^{1/2}\Big >}
+\nonumber \\
+\frac{\Big < \Big [ \sum_j \Big ( \frac{{n_j} S_{n_j}}{S_1}
\Big )^{1/2} \Big ]^2 \Big >}
{\Big < \sum_j \Big ( \frac{{n_j} S_{n_j}}{S_1}
\Big )^{1/2}\Big >}  f_2^2
-
\frac{  \Big < \sum_j {n_j}^{3/4} \Big ( \frac{S_{n_j}}{S_1} \Big )^{1/4}
\sum_j \Big ( \frac{{n_j} S_{n_j}}{S_1} \Big )^{1/2} \Big >}
{\Big < \sum_j \Big ( \frac{{n_j} S_{n_j}}{S_1}
\Big )^{1/2}\Big >} 2 f_2 \Big ] \mu_1 <p_T>_1^2 \nonumber \\
=[ f_3 +f_4 f_2^2 -2 f_2 f_5] \mu_1 <p_T>_1^2 \ .
\eea
Finally we arrive to:
\beq
F_{p_T}= \sqrt{\mu_1} \sqrt{\frac{f_3+f_4 f_2^2 -2 f_2 f_5}{f_1-f_2^2}} - 1\ .
\eeq
where $f_1$, $f_2$, $f_3$, $f_4$ and
$f_5$ are defined in the expressions (9)-(11).

In order to compute eq. (12), several ingredients are necessaries. On one hand
we need a Monte Carlo code \cite{21}-\cite{22}
for the cluster formation, in order to compute the
number of strings that come into each cluster and the area of the cluster. On the other hand, 
we do not use a Monte
Carlo code for the decay of the cluster, since we apply analytical expressions (eqs. (4)) for the
transverse momentum and the multiplicities of the clusters.
 
We also need  
the value of $\mu_1$  --multiplicity produced by one individual string--. 
It was previously fixed from a comparison of the model to SPS and RHIC data \cite{19}-\cite{20}
on multiplicities. In the first Ref. of \cite{19}, the total
multiplicity per unit
rapidity 
produced by one string has been taken as $\mu_{0\, {tot}} \simeq 1$. If we assume that
$2/3$ of the created particles are charged, that would lead to a charged
particle multiplicity per unit rapidity for each individual string of
$\mu_{0\, {ch}}=0.65$. 
In order to compare with experimental data we define $\mu_1=\mu_{0\, {ch}}\, y$, 
where
$y$ is the rapidity interval of the produced particles. 
We don't introduce any dependence of $\mu_0$ with the energy or the centrality
of the collision.
Notice that in (12) the value of $<p_T>_1$ cancels.

We have neglected the subsequent rescattering of hadrons and
resonances that takes place after the decay of the clusters. It gives rise to
correlations 
which would be similar in the clustering approach
and in an independent string picture, unless additional dynamics were taken into account. 
Therefore its contribution to $F_{p_T}$
cancels in our approach. 

The comparison of our results for the dependence of $F_{p_T}$
on the number of
participants $N_p$ with the PHENIX data \cite{15b} is shown in Fig. 1. 
The calculation is done for 
charged particles in the rapidity 
range $|\eta| < 0.35$, $\mu_1= 0.7\, \mu_{0\, {ch}}$.
An acceptable overall
agreement is obtained.

In order to compute our value for
$F_{p_T}$, we take into account all possible transverse momenta, whereas in
the experiment there is a limited acceptance, 0.2 GeV/c $< p_T < p_T^{max}$. 
PHENIX \cite{15}-\cite{15b} has studied the variation 
of $F_{p_T}$ with the maximal value of the acceptance for $p_T$, $p_T^{max}$.
The maximum of $F_{p_T}$ is reached for the largest acceptance, $p_T^{max}=4$
GeV/c \cite{15}. So 
we can expect that our value for $F_{p_T}$ is going to be higher than the experimental one,
specially for a moderate number of participants,
$N_p$, 
since the truncated average $p_T$ \cite{0207009}, 
$<p_T^{trunc}>= \frac{\int_{p_T^{min}}^{\infty} p_T
dN/dp_T}{\int_{p_T^{min}}^{\infty} dN/dp_T} -p_T^{min}$, decreases with the number of participants
for $p_T^{min} > 2$ GeV/c. This means that,
for momenta higher than 2 GeV/c, the high $p_T$ contribution would be due to collisions with a moderate number
of participants.
%
These considerations
may explain the difference between our results and PHENIX data --with a limited acceptance
of 0.2 GeV/c $< p_T < 2.0$ GeV/c-- at low $N_p$.

In Fig. 2 our results for $\phi_{p_T}$ of charged particles in Pb-Pb central
collisions at 158 AGeV are compared with the experimental data of NA49
Collaboration \cite{23a}. In this case  the data correspond to the forward
rapidity range $4.0 
< y < 5.5$. For this reason we use 
$\mu_1= 1.5\, \mu_{0\, {ch}}$, which in principal implicates larger
correlations.  However we see that this effect is compensated, since we have a lower 
value for
the mean number 
of strings at fixed $N_p$, due to: 

\noindent
a) lower energy at SPS than at RHIC so less strings are produced,

\noindent
b) the mean number of strings in this rapidity region is proportional to the
number of participants $N_p$, while in the central region it is proportional to
$N_p^{4/3}$ due to the contribution of $q-\bar{q}$ strings from the sea. 

For the computation of $\phi_{p_T}$ we use $<p_T>_1=0.3$ GeV/c.

The CERES Collaboration has also measured $\phi_{p_T}$ \cite{23a} at four
different centralities: $0-5 \%$, $5-10 \%$, $10-15 \%$ and $15-20 \%$ for Pb-Au
collisions at 40, 80 and 158 AGeV/c in the central pseudorapidity range $2.2 <
\eta < 2.7$ and restricted to tracks with transverse momenta $0.1 < p_T < 1.5$
GeV/c. Due to the narrower rapidity and transverse momentum range one would expect
a lower $\phi_{p_T}$ value. However this is compensated with the increase
of the number of strings  at central
rapidities. The data at 158 AGeV, 
after short range removal, are 3.3, 3.6, 4.4 and
4.1 for the above mentioned centralities, with errors of the order of 1.5
--for smaller
energies the data are lower as expected--.
These values and their dependence with centrality are compatible with our results
of Fig. 2.

In order to have a better understanding of the behaviour of $F_{p_T}$ and
$\phi_{p_T}$ on the number of participants, we plot in Fig. 3 and Fig. 4
the mean number of clusters $M$ and the dispersion on the number of clusters
multiplied by the number of clusters
$\sigma_M * M$ at RHIC and SPS energies.
The ratio $\sigma_M^2 / M$ would be one in the case of a Poisson distribution. 
The $p_T$ fluctuations are due in our approach to the different mean transverse
momenta of the clusters. These momenta depend on the number of strings that comes
into the cluster and the area occupied by the cluster through our eq. (4),
therefore $M$ and $\sigma_M$ should be the key quantities. 
However, $\sigma_M^2 / M$ ranges between $1/2$ and 2 in the whole $N_p$ range, what 
indicates a lower variation than the one for $M$, as can be seen from Fig. 3 and 4
where $M$ and $\sigma_M * M$ are plotted. The only effect of $\sigma_M$ is 
to shift the maximum of $M$.
Because
of this we expect the dependence of $F_{p_T}$ and
$\phi_{p_T}$ on $N_p$ to be more similar to the $M$ behaviour, as it is actually.
In other words, a decrease in the number of effective sources leads to a decrease of the tranverse momentum
fluctuations.

Similar conclusions have been reached in Ref. \cite{dias}, where a formula for $F_{p_T}$
as a function of the cluster dispersion over the mean number of strings per
cluster has been obtained.

Notice that we only need to know the number of
strings formed for each centrality and their location in the impact parameter
space in order to form clusters. This information, together with eq. (4), is
enough for us to calculate $F_{p_T}$ and
$\phi_{p_T}$. The same variables have been able to
describe the behaviour of the strength of two \cite{23} and three \cite{24} body Bose-Einstein
correlations with centrality and the dependence of the multiplicities and
transverse momentum distributions \cite{20}, \cite{26} on the centrality. All that points out
that 
the percolation approach may be appropriate to describe 
the relativistic heavy ion collisions.

We thank J. Dias de Deus for useful discussions.
This work has been done under Contract No FPA2002-01161 from CICYT of Spain and
FEDER from EU.

\begin{figure}
\centering\leavevmode

\vskip -1cm
\epsfxsize=7in\epsfysize=7in\epsffile{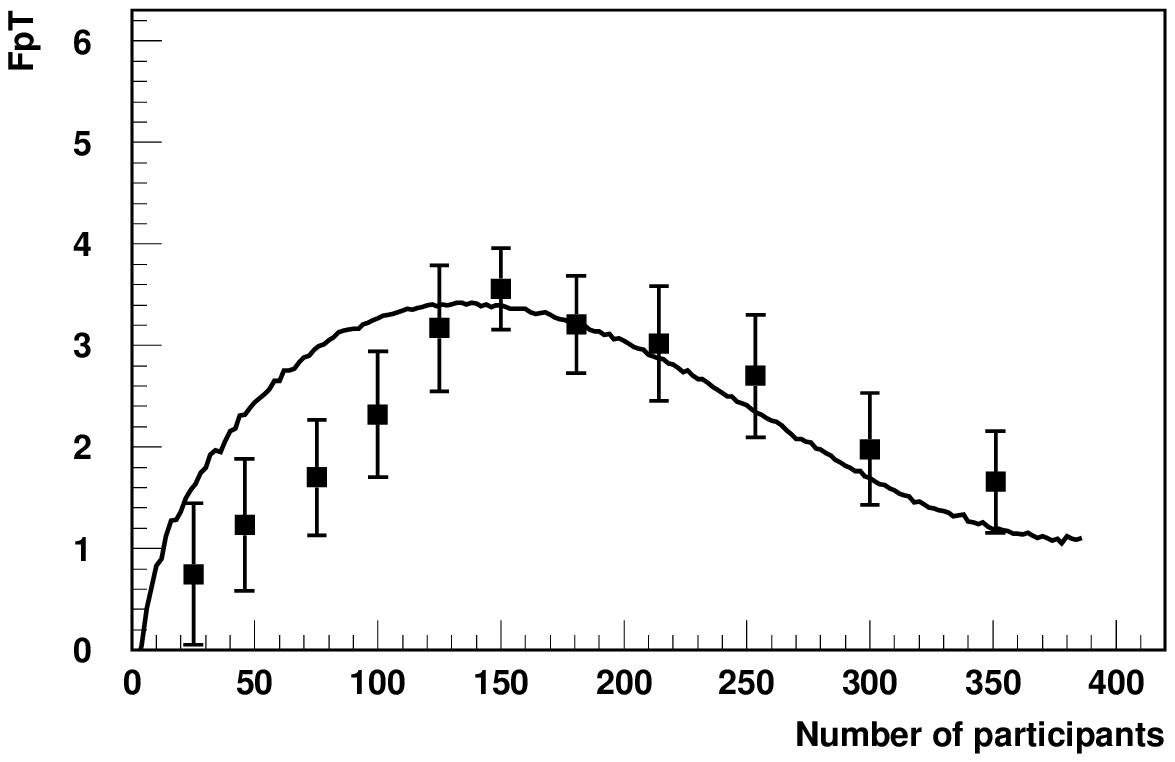}
\vskip 0.5cm
\caption{$F_{p_T} (\%)$ 
versus the number of participants. Experimental
data from PHENIX at $\sqrt{s}=200$ GeV are compared with our results
(solid line).}
\label{figure1}
\end{figure}

\begin{figure}
\centering\leavevmode
\epsfxsize=6in\epsfysize=6in\epsffile{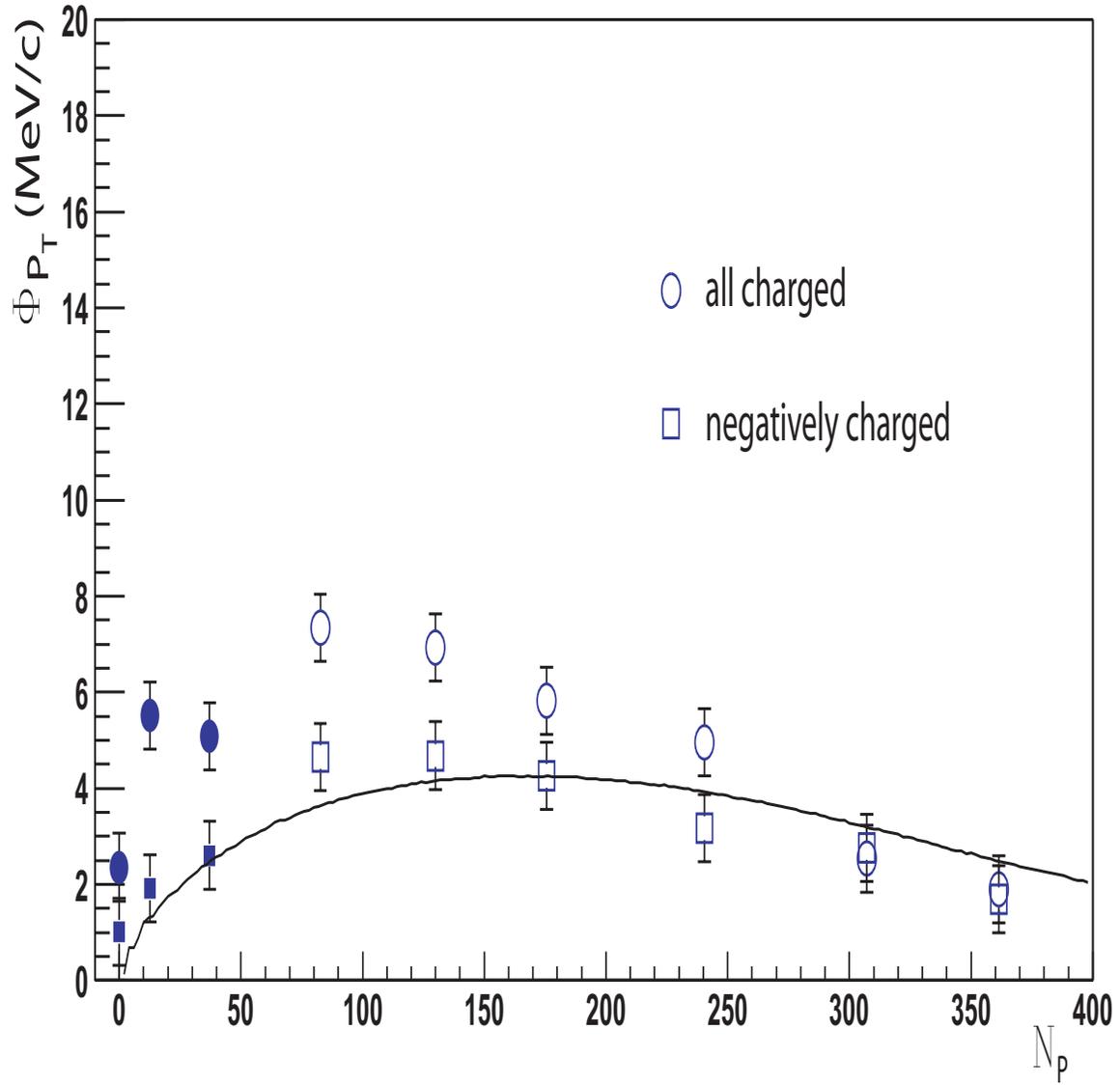 }
\vskip 0.5cm
\caption{
$\phi_{p_T}$ versus the number of participants.
Experimental data from NA49
Collaboration at SPS energies are compared with our results
(solid line).}
\label{figure2}
\end{figure} 

\begin{figure}
\centering\leavevmode
\epsfxsize=7in\epsfysize=7in\epsffile{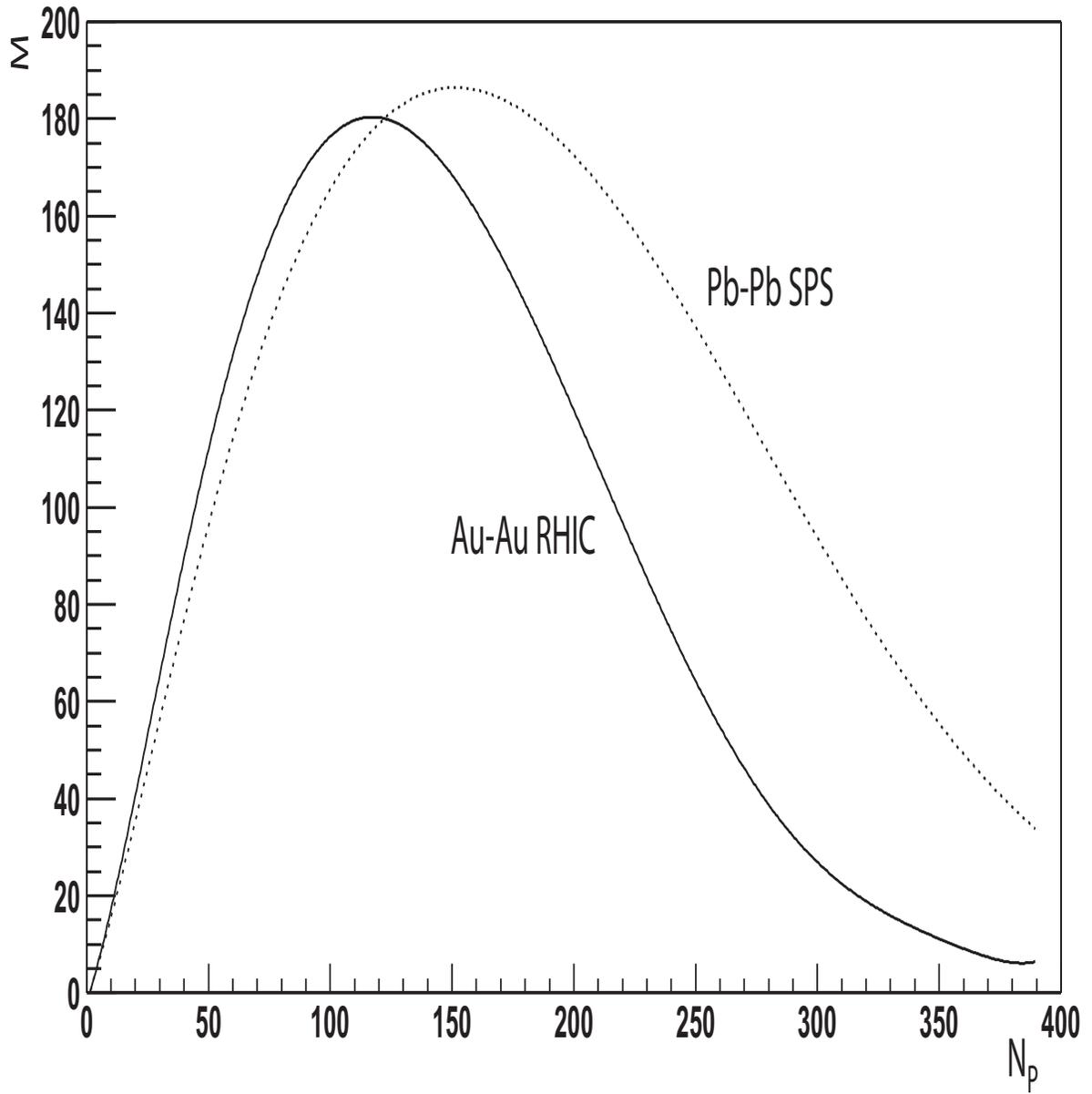}
\vskip 0.5cm
\caption{
Mean number of clusters $M$ versus the number of participants
for Pb-Pb collisions at SPS energies (dotted line) and Au-Au collisions
at RHIC energies (solid line).} 
\label{figure3}
\end{figure} 

\begin{figure}
\centering\leavevmode
\epsfxsize=6in\epsfysize=6in\epsffile{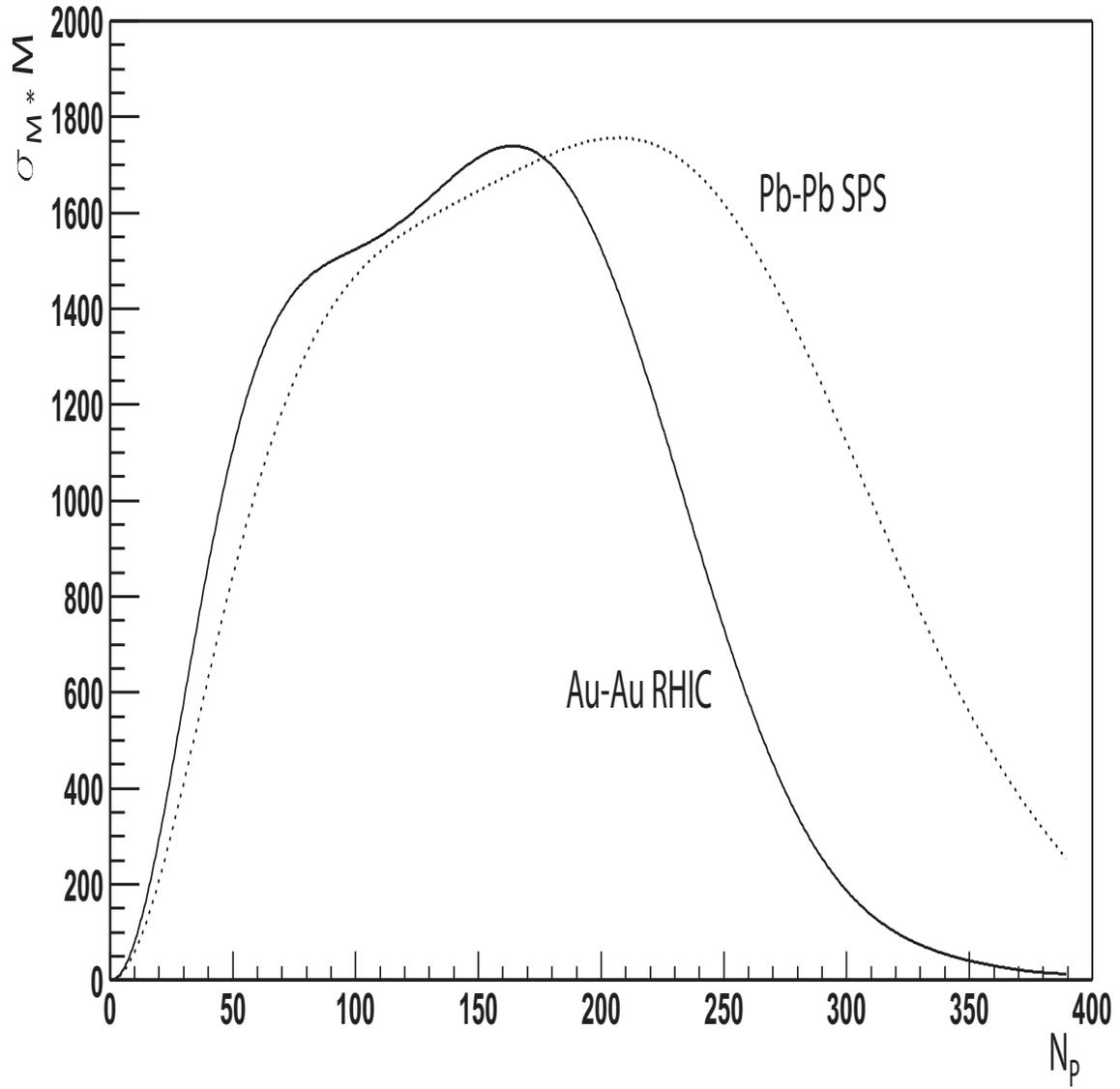}
\vskip 0.5cm
\caption{
Dispersion on the number of clusters 
multiplied by the number of clusters $\sigma_M * M$ 
versus the number of participants
for Pb-Pb collisions at SPS energies (dotted line) and Au-Au collisions
at RHIC energies (solid line).} 
\label{figure4}
\end{figure}

\end{document}